\documentclass[12pt]{iopart}
\usepackage{graphicx}
\usepackage{psfig}
\usepackage{wrapfig}
\usepackage{epsfig}
\begin{document}

\title[PHENIX Heavy Flavor]{The PHENIX measurement of Heavy Flavor via
  Single Electrons in pp, d-Au and Au-Au collisions at $\bf{\sqrt{s_{NN}} = 200~GeV}$}

\author{Sean Kelly for the PHENIX Collaboration
\footnote[3]{For the full PHENIX Collaboration author list and
acknowledgments, see Appendix "Collaborations" of this volume.}  }

\address{University of Colorado, Boulder}

\begin{abstract}

The PHENIX experiment at RHIC has measured single electron spectra in
proton-proton, deuteron-gold and gold-gold collisions at
$\sqrt{s_{NN}} = 200~GeV$.  The photonic contribution (photon
conversions and Dalitz decays) is subtracted from the inclusive
spectra resulting in non-photonic single electron spectra.  The
principal source of non-photonic electrons is the
semi-leptonic decay of charm and bottom mesons.  The implications for
heavy flavor production in hot and cold nuclear systems are discussed.
 
\end{abstract}




\section{Introduction}

Heavy flavor production in nuclear systems has utility as a probe of both
cold nuclear matter effects
and of more exotic effects associated with the
hot partonic state formed in the aftermath of high energy
nucleus-nucleus collisions.  
Differences in the relative yield of heavy
flavor in $p-p$ and $p-A$ collision probes nuclear modifications of the
gluon distribution function and energy loss in a cold nuclear medium.
Strong suppression of high $p_{T}$ hadrons of light quark flavor
at moderate to high transverse momentum has been
observed at RHIC~\cite{Adler:2003qi}.
Medium induced gluon radiation, which is considered as the main
cause of the high $p_{T}$ suppression, may be reduced for heavy
quarks~\cite{Djo} \cite{Dok}.

\section{Data Analysis and Results}

PHENIX measures non-photonic electron spectra by first measuring the
inclusive electron spectra and then subracting the contribution from
photonic sources.  The principal contributors to the photonic
electrons are Dalitz decays of $\pi^{0}$ and $\eta$ mesons and photon
conversions, and the dominant contributor to the non-photonic electrons
is heavy quark decay.  Other non-photonics
contributors that enter at the percent level are
light vector meson di-electron decay and
$K_{e3}$.  The inclusive electron spectra are measured with a combination
of detectors in the PHENIX central arm spectrometers
($|\eta| < 0.35$ in pseudorapidity) which is
described in detail elsewhere \cite{Ham02}.  The momentum of charged
tracks is reconstructed using drift and pad chambers.  Electron
identification is accomplished through a combination of an associated
signal in a ring imaging Cerenkov detector (RICH) and an $E/p$ cut
that employs an electomagnetic calorimeter (EMCAL) for the energy
measurement.

\begin{figure}[th]
\begin{center}
\begin{minipage}{\textwidth}
\includegraphics[width=0.5\textwidth]{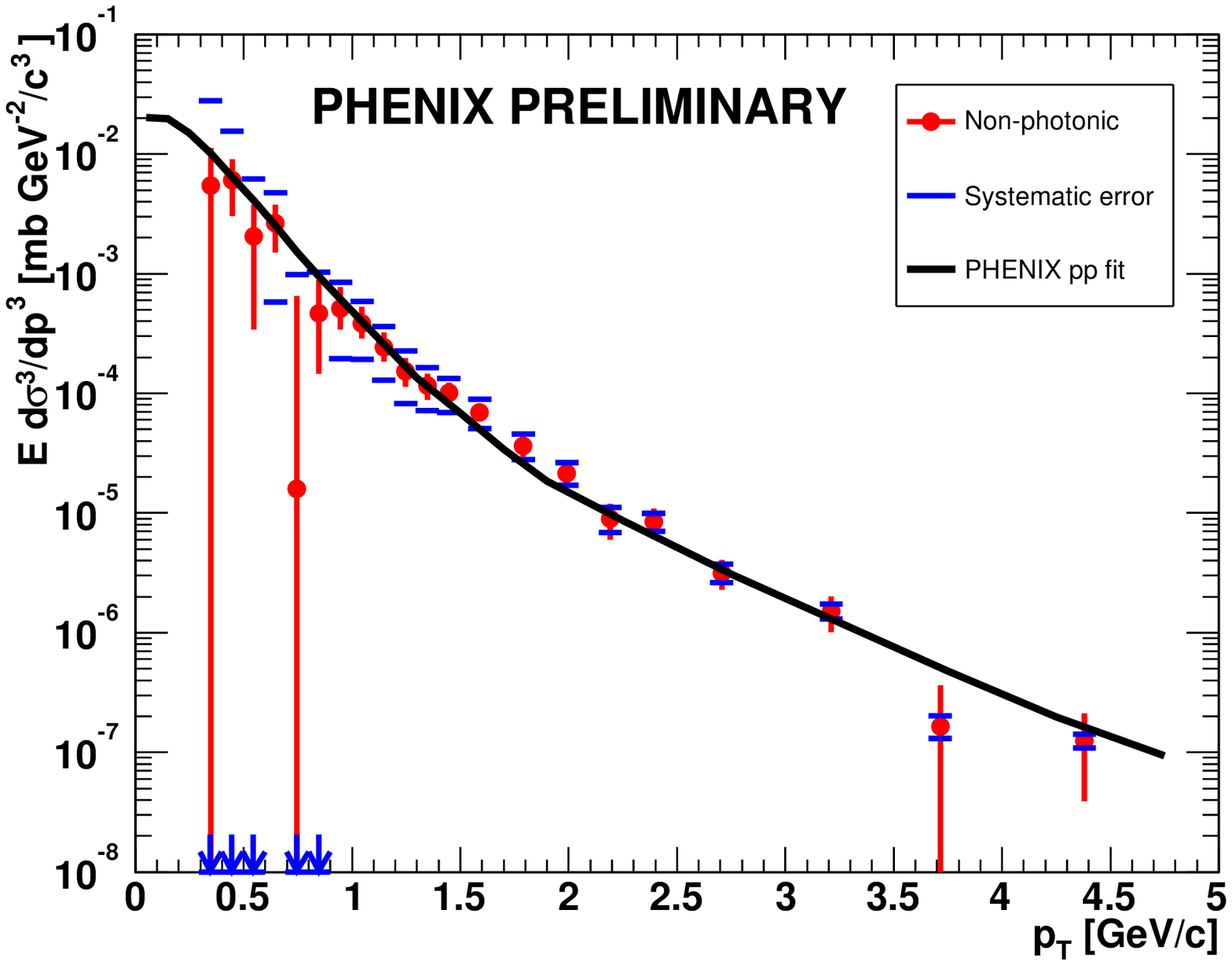}
\includegraphics[width=0.5\textwidth]{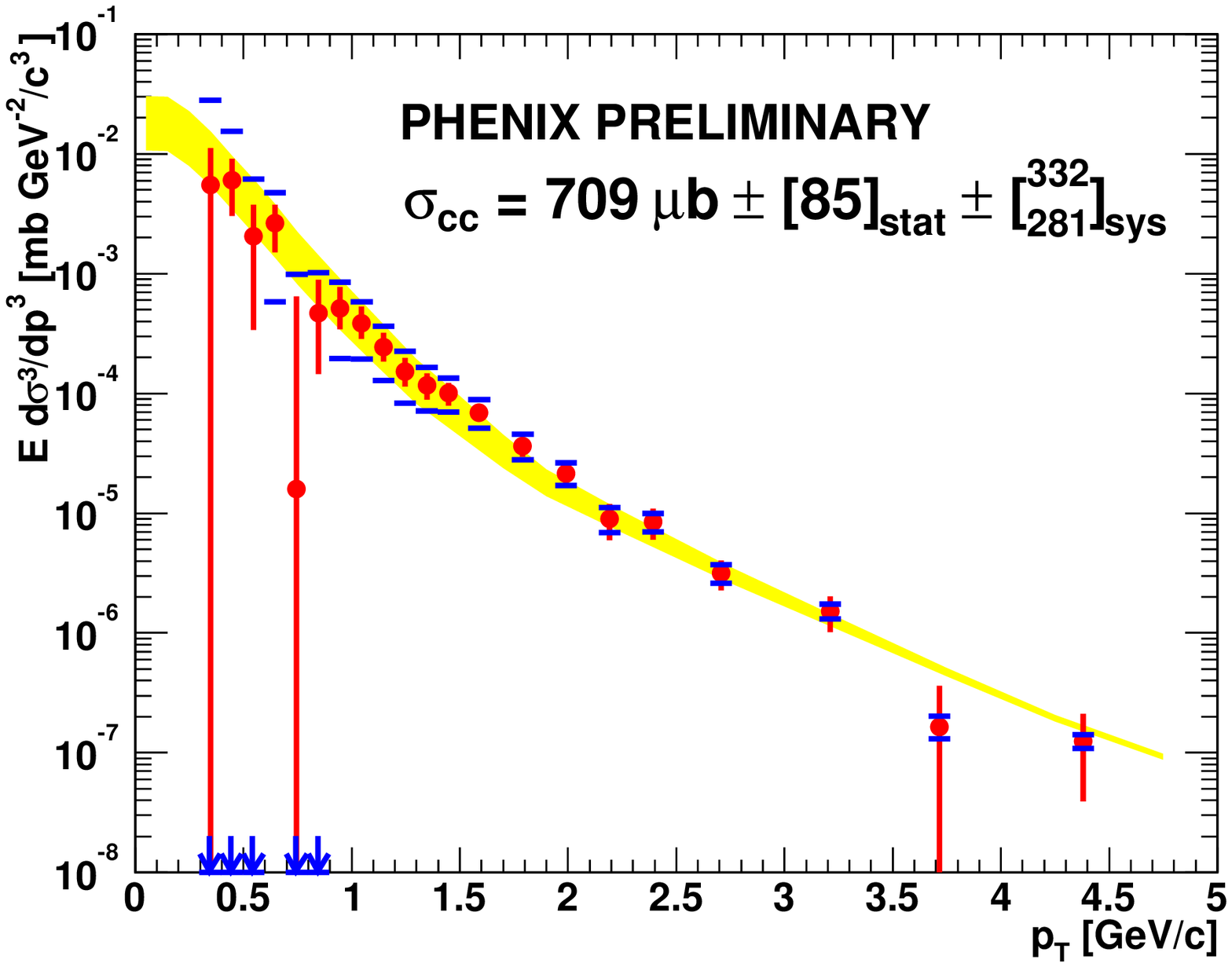}
\end{minipage}
\caption{
Right and left panels show the same non-photonic 
single electron spectra from $p-p$ collisions at $\sqrt{s_{NN}} = 200~GeV$ 
with an empirical liine shape fit (left) and the error band for determining the 
charm cross section (right).}
\label{fig:pp_with_fit}
\end{center}
\end{figure}

The dataset used for the proton-proton analysis consisted of 15M
min-bias triggers and 420M sampled min-bias triggers associated with a
level 1 trigger that required a coincidence between the EMCAL and RICH 
(ERT).  The data have been corrected for acceptance and reconstruction
efficiency, and the photonic background has been subtracted using a
cocktail method that is described in detail elsewhere \cite{Adcox}.
Fig.~\ref{fig:pp_with_fit} shows the spectrum of non-photonic
electrons from $p-p$ collisions.  A fit to the data is shown as the solid
line.  The right panel shows the total charm cross section determined
by allowing PYTHIA lineshapes for electrons from charm and bottom
meson decay to assume normalizations that best fit the data.  The
systematic error was calculated by offseting the data by the
systematic errors and repeating the fit proceedure.  The total charm
cross section derived using this extrapolation technique is
$\sigma_{cc} = 709\mu b \pm[85]_{stat} \pm [^{332}_{281}]_{sys}$.

\begin{figure}
\begin{center}
\epsfig{figure=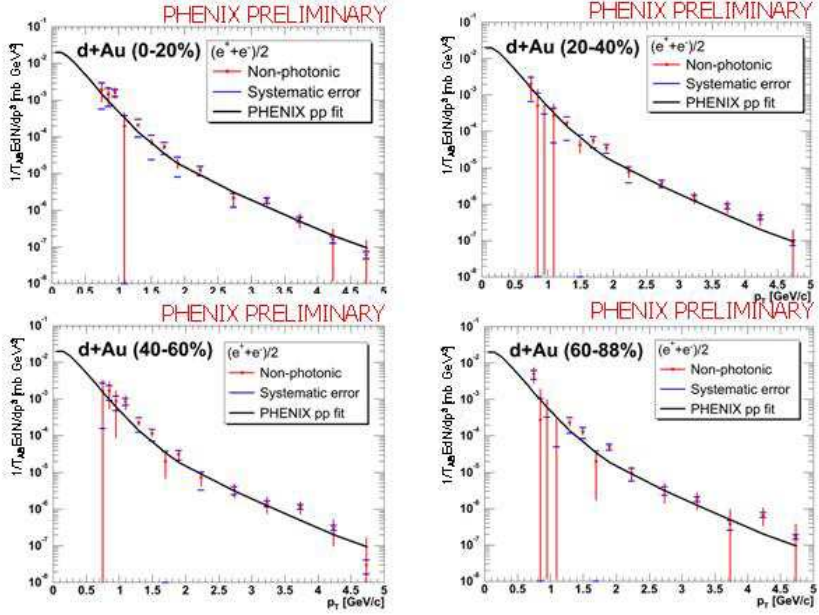, width=11cm}
\caption{Non-photonic single electron spectra from $d-Au$ $\sqrt{s_{NN}} = 200~GeV$}
\label{fig:dau_centrality}
\end{center}
\end{figure}

The method by which electrons from photonic sources were subtracted in
the analysis of the $d-Au$ data from Run 3 at RHIC was the so-called
converter subtraction method.  During a fraction of the running period
a photon converter of known mass was inserted close to the interaction
point.  By measuring the electron yield with and without the converter
it is possible to infer the fraction of measured electrons from photonic
sources.  The preliminary analysis of the $d-Au$ data used a converter
run sample that consisted of 5M min-bias events and the equivalent of
312M min-bias events sampled by the ERT trigger.  The non-converter
dataset included 5M min-bias and 600M trigger sampled min-bias
events. Fig.~\ref{fig:dau_centrality} shows the non-photonic single
electron spectra from deuteron-gold collisions at $\sqrt{s_{NN}} = 200~GeV$ 
in four centrality classes scaled down by the nuclear thickness.  
The solid line is the best fit to $p-p$ data.

\begin{figure}
\begin{center}
\epsfig{figure=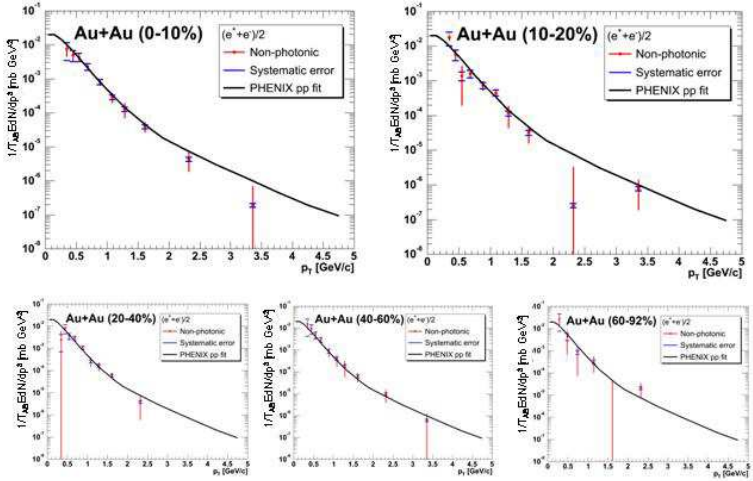, width=10cm}
\caption{Non-photonic single electron spectra from Au-Au $\sqrt{s_{NN}} = 200~GeV$}
\label{fig:auau_centrality}
\end{center}
\end{figure}

\begin{figure} [th]
\begin{center}
\begin{minipage}{\textwidth}
\includegraphics[width=0.5\textwidth]{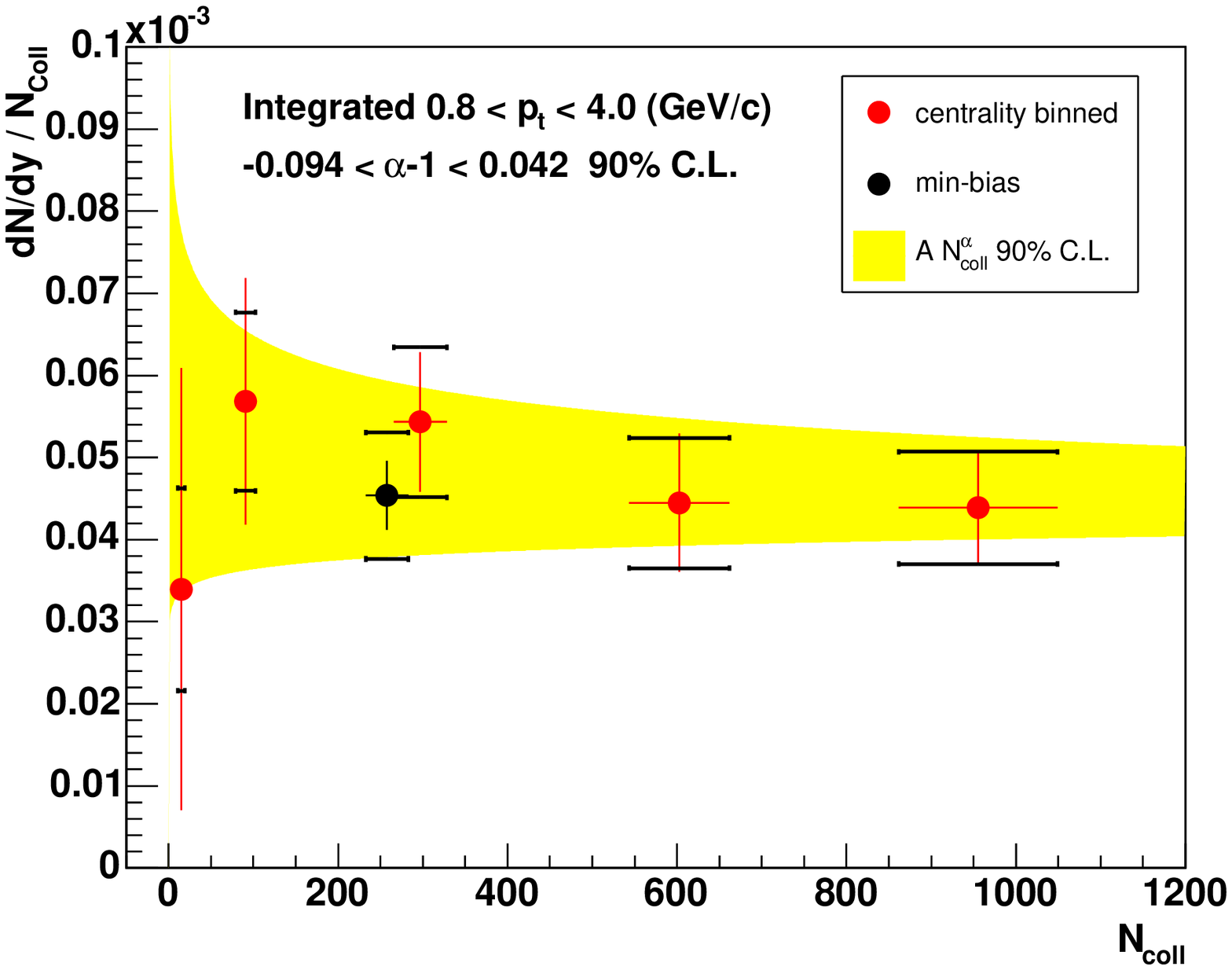}
\includegraphics[width=0.5\textwidth]{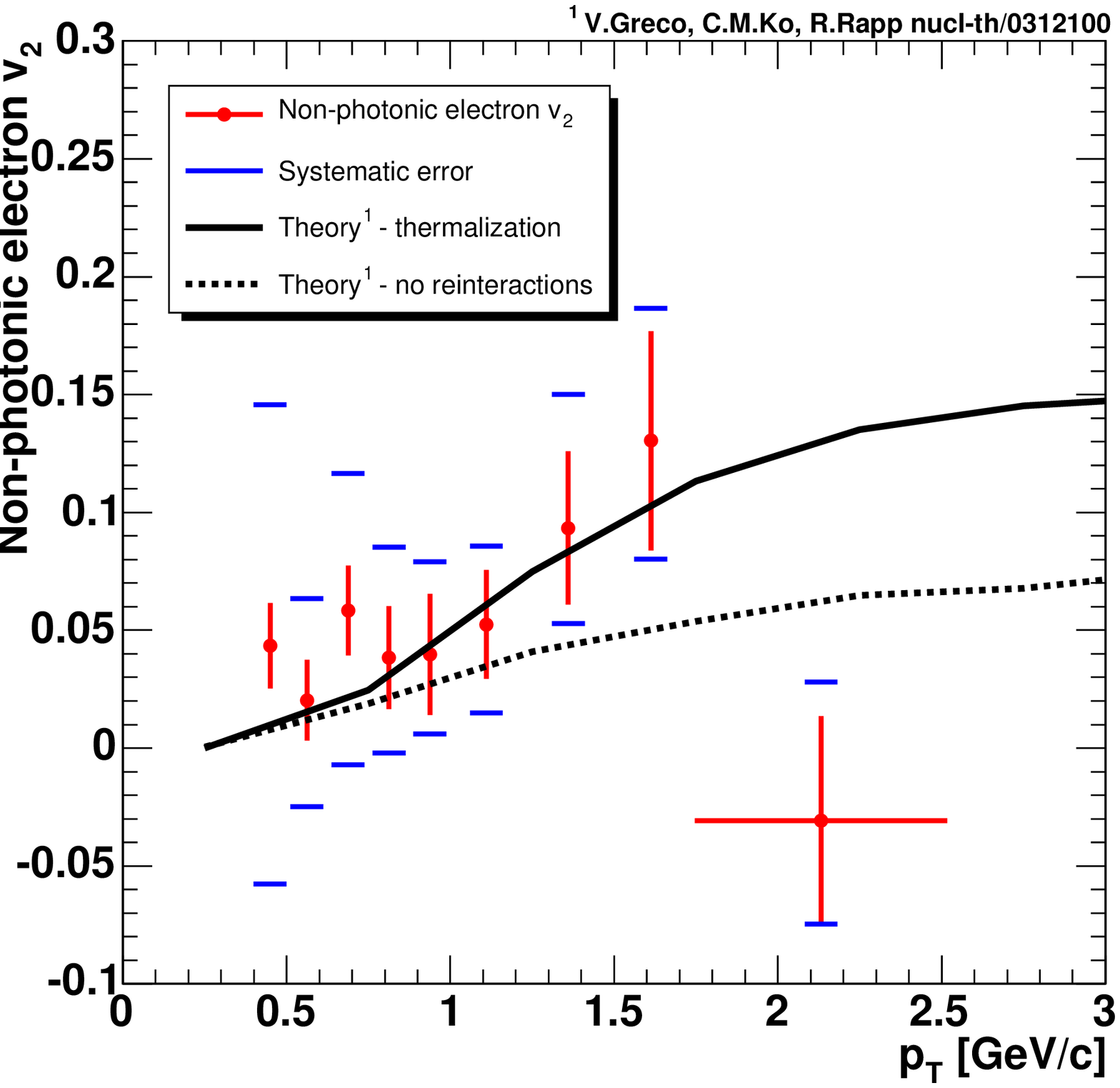}
\end{minipage}
\caption{(Left) PHENIX Preliminary integrated $dN/dy/N_{binary}$ versus $N_{binary}$.  
(Right) PHENIX Preliminary Non-photonic  electron $v_{2}$}
\label{fig:auau_dndy}
\end{center}
\end{figure}

The analysis of $Au-Au$ collision data for single electrons also
employed the converter subtraction method.  These data 
consist of 2.2M min-bias and 2.5M min-bias
events for the converter and non-converter runs respectively.
Fig.~\ref{fig:auau_centrality} shows the non-photonic single electron
spectra in five centrality bins with the best fit to the $p-p$ data
overlayed.  The extent to which the single electron yield in $Au-Au$
collision is consistent with a binary scaling hypothesis is shown in
Fig.~\ref{fig:auau_dndy}.  

\section{Conclusion}

Within errors, the $d-Au$ non-photonic single electron spectra are in
good agreement with the $p-p$ spectra in all centrality classes.  This
is indicative of no strong modification to the initial state gluon
distribution function in the $x$ range relevant to heavy quark
production at central rapidity $(10^{-2} < x < 10^{-1})$.  The
integrated $dN/dy$ in $Au-Au$ collisions is consistent with a binary
scaling hypothesis, indicating no strong charm enhancement 
or suppression in heavy
ion collisions.  No conclusion can be drawn with regards to a
modification of the spectra shapes in $Au-Au$ collision, as to whether
it follows pQCD, exhibits significant charm quark energy loss, or
follows hydrodynamic flow.
Measuring $v_{2}$ of charm was proposed \cite{Batsouli} as a potential
way to disambiguate these very different dynamical
scenarios. Figure ~\ref{fig:auau_dndy} shows the preliminary measurment of
non-photonic single electron flow in $Au-Au$ collision.  The current
$Au-Au$ run at RHIC will provide the necessary statistics to investigate
both heavy quark energy loss and non-photonic electron flow.\linebreak


\begin{thebibliography}{99}

\bibitem{Adler:2003qi} S.~S.~Adler {\it et al.},Phys.\ Rev.\ Lett.\  {\bf 91}, 072301 (2003).

\bibitem{Djo} M. Djordjevic, M. Gyulassy Nucl.Phys.{\bf A733}, 265-298 (2004).

\bibitem{Dok}  Y.L. Dokshitzer and D.E. Kharzeev, Phys. Lett.{\bf B519}, 199 (2001).

\bibitem{Ham02}  H. Hamagaki {\it et al.} (PHENIX Collaboration), Nucl. Phys.{\bf A698}, 412 (2002).

\bibitem{Adcox} K. Adcox {\it et al.} Phys. Rev. Lett. {\bf 88},192303 (2002).

\bibitem{Batsouli} S. Batsouli, S. Kelly, M. Gyulassy, J.L. Nagle, Phys.Lett. {\bf B557} 26-32 (2003) 

\end{thebibliography}
\end{document}